\documentclass[a4paper]{article}
\usepackage{amsmath}
\usepackage{amssymb}
\usepackage{graphicx}
\usepackage{epstopdf}

\def\ii{\int\limits_{\mathbb{R}}}
\def\jj{\int\limits_{-h}^{\eta(x,t)}}

\def\jjj{\int\limits_{\eta(x,t)}^{h_1}}

\begin{document}
\section*{\LARGE \bf{The Dynamics of Flat Surface Internal Geophysical Waves with Currents}}

\begin{center}

{\large \bf Alan Compelli$^{a,\dag}$} and {\large \bf  Rossen I.
Ivanov$^{a,\ddag}$} 

\vskip1cm

\hskip-.3cm
\begin{tabular}{c}
\\
$\phantom{R^R}^{a}${\it School of Mathematical Sciences, Dublin Institute of Technology, }\\ {\it Kevin Street, Dublin 8, Ireland} \\
\\
\\
\\{\it $^\dag$e-mail: alan.compelli@dit.ie}
\\{\it $^\ddag$e-mail: rossen.ivanov@dit.ie  }
\\
\hskip-.8cm
\end{tabular}
\vskip1cm
\end{center}
\input epsf

\begin{abstract}
\noindent{\sc }
A two-dimensional water wave system is examined consisting of two discrete incompressible fluid domains separated by a free common interface. In a geophysical context this is a model of an internal wave, formed at a pycnocline or thermocline in the ocean. The system is considered as being bounded at the bottom and top by a flatbed and wave-free surface respectively. A current profile with depth-dependent currents in each domain is considered. The Hamiltonian of the system is determined and expressed in terms of canonical wave-related variables. Limiting behaviour is examined and compared to that of other known models. The linearised equations as well as long-wave approximations are presented.
\\
\\
{\bf Mathematics Subject Classification (2010):} Primary 35Q35; Secondary 37K05, 74J30
\\
\\
{\bf Keywords:} Internal waves, Equatorial undercurrent, shear flow, Hamiltonian system, KdV equation

\end{abstract}

\section{Introduction}

The addition of current considerations to water wave systems introduces various degrees of complexity depending on the current profile under study. Understanding the interaction of waves and currents provides applications for the prediction of tsunami, rogue waves, etc. and is of interest to oceanographers and climatologists amongst many interested groups.

For a general description of the problem of waves and currents we refer to the following reviews and monographs \cite{ACbook, Per, IGJ, TK} and the references therein.

For ocean waves of large magnitude, the viscosity does not
play an essential role and can be neglected, so effectively the fluid dynamics are given by Euler's equations.
In 1968 V.E. Zakharov in his work \cite{Zak} demonstrated that irrotational waves on deep water have a canonical Hamiltonian formulation (i.e. zero vorticity and one layer). The result has been extended to models with finite depth and shear with constant vorticity \cite{NearlyHamiltonian, Wahlen2}. The study of such essentially non-linear waves and their interaction with currents is of utmost importance in the advancement of geophysical fluid dynamics. Most predictions of water-wave propagation in oceanography use linear approximations. Although this approach is successful in many instances, for complex flow patterns an adequate description of the phenomenon can not neglect nonlinear effects. For this reason we aim to develop a nonlinear approach that captures the main features of the dynamics and is also amenable to approximations (linear or weakly nonlinear) in specific physical regimes, thus enabling an in-depth study in these circumstances.

Several irrotational \cite{Zak,BenjOlv, Craig1,Milder,Miles0,Miles} and rotational models \cite{Constantin1,Constantin2,ConstantinSattingerStrauss,ConstantinStrauss, ConstantinEscher1, TelesdaSilva, NearlyHamiltonian, Wahlen2} and in particular the model of 2-media systems with internal waves such as \cite{BenjBrid, BenjBridPart2, Craig1, Craig2, Compelli, Compelli2,M} form the framework for the study being undertaken. The consideration of wave-current interactions has been explored in several recent studies \cite{CJ,CI,CompelliIvanov1,CompelliIvanov2,CHP}.

This paper considers a two-media system where the media are separated by an internal wave. The bottom of the system is bounded by an impermeable flat-bed. The top of the system is also considered to be a flat surface. It is important to note that, by comparison to \cite{ConstIvMartin}, this is not equivalent to assuming the amplitude of the surface waves as being small. One has to keep in mind that no matter how small the surface waves are there is a coupling between surface and internal waves. 

In geophysical models, for example in the equatorial region, the Pacific is characterised by a thin shallow layer of warm and less dense water over a much deeper layer of cold denser water. The two layers are separated by a sharp thermocline (where the temperature gradient has a maximum, it
is very close to the pycnocline, where the pressure gradient has a maximum) at a depth, depending on the location, but usually at 100--200m beneath the surface. Both layers are homogeneous and their sharp boundary is the thermocline/pycnocline. Internal waves are formed at this boundary. 

This paper aims to generalise the studies of internal waves by considering different vorticities and current profiles in each medium. Recovery of already known special cases will be demonstrated.

\section{Setup and governing equations}
\begin{figure} 
\centering
\includegraphics[width=0.7\textwidth]{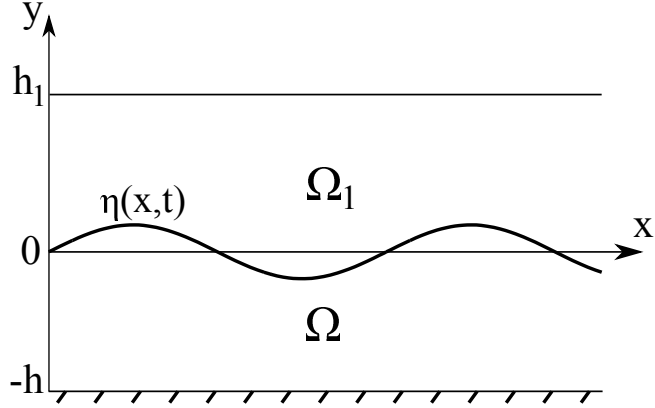}
\caption{System setup. The function $\eta(x,t)$ describes the elevation of the internal wave.}
\label{fig:1}
\end{figure}

A two-dimensional water wave system consisting of two discrete fluid domains separated by a free common interface in the form of an internal wave, such as a pycnocline or thermocline, is studied as per Figure \ref{fig:1}. 

The system is bounded at the bottom by an impermeable flatbed and is considered as being bounded on the surface by an assumption of absence of surface motion. The domains $\Omega=\{(x, y)\in\mathbb{R}^2: -h< y < \eta(x,t)\}$ and $\Omega_1=\{(x, y)\in\mathbb{R}^2: \eta(x,t)< y < h_1\}$ are defined with values associated with each domain using corresponding respective subscript notation. Also, subscript $c$ (implying \emph {common interface}) will be used to denote evaluation on the internal wave. Propagation of the internal wave is assumed to be in the positive $x$-direction which is considered to be 'eastward'. The centre of gravity is in the negative $y$-direction. 

The function $\eta(x,t)$ describes the elevation of the internal wave with the mean of $\eta$ assumed to be zero, $$\int_{\mathbb{R}} \eta(x,t) dx=0.$$ 

The system is considered incompressible with $\rho$ and $\rho_1$ being the respective constant densities of the lower and upper media and stability is given by the immiscibility condition
\begin{alignat}{2}
\label{stability}
\rho>\rho_1.
\end{alignat}

The velocity fields ${\bf{V}}(x,y,t)=(u,v)$ and ${\bf{V}}_1(x,y,t)=(u_1,v_1)$ of the lower and upper media respectively are defined in terms of the respective velocity potentials 
\begin{equation} \label{phidec}
        \left\lbrace
        \begin{array}{lcl}
        \varphi \equiv \tilde{\varphi}+\kappa x \mbox{ for $\Omega$}
        \\
       \varphi_1 \equiv\tilde{\varphi}_1+\kappa_1 x \mbox{ for $\Omega_1$}
        \end{array}
        \right.
\end{equation}
and stream functions  $\psi$ and $\psi_1$  as
\begin{equation}
\label{def__stream_phi}
        \left\lbrace
        \begin{array}{lcl}
  \left.        \begin{array}{lcl}
        u  =   \tilde{\varphi}_{x} +\gamma y+\kappa=\psi_y
        \\
        v=  { \tilde{\varphi}}_{y}=-\psi_x
        \end{array}\right\} \mbox{for $\Omega$}
\\
  \left.        \begin{array}{lcl}
        u_1  =   \tilde{\varphi}_{1,x} +\gamma_1 y+ \kappa_1=\psi_{1,y}
        \\
        v_1=  { \tilde{\varphi}}_{1,y}=-\psi_{1,y}
        \end{array}\right\} \mbox{for $\Omega_1$}
        \end{array}
        \right.
\end{equation}
where $\gamma=u_y-v_x$ and $\gamma_1=u_{1,y}-v_{1,x}$ are the constant non-zero vorticities.

The motivation for the decomposition \eqref{phidec} is the separation of the velocity potential to  ``wave-motion'' components, given by $\tilde{\varphi}$ and $\tilde{\varphi}_1$, and the $\kappa x$ and $\kappa_1 x$ terms, generating constant horizontal velocity components of the flows in the corresponding domains.

We make the assumption that the functions $\eta(x, t),$ $\tilde{\varphi}(x, y, t)$ and $\tilde{\varphi}_1(x, y, t)$ belong to the Schwartz class $\mathcal{S}(\mathbb{R})$ with respect to the $x$ variable (for any $y$ and $t$). This reflects the localised nature of the wave disturbances. The assumption of course implies that for large absolute values of $x$ the internal wave attenuates \begin{equation}
\lim_{|x|\rightarrow \infty}\eta(x,t)=0, \quad \lim_{|x|\rightarrow \infty}{ \tilde{\varphi}}(x,y,t)=0 \quad\mbox{and} \quad \lim_{|x|\rightarrow \infty}{ \tilde{\varphi}_1}(x,y,t)=0.
\end{equation}

Effectively  the current profiles in $\Omega$  and $\Omega_1$ are $U(y)=\gamma y+\kappa$ and $U_1(y)=\gamma_1 y+\kappa_1$, correspondingly. Here $\kappa$ and $\kappa_1$ are constants, representing the average flow in each respective domain, see \cite{ConstIvMartin}, where the situation with a free surface is studied. 

Considering equatorial motion (equatorial current and undercurrent coupled to internal waves) the following Coriolis forces per unit mass have to be taken into account:
\begin{equation}
        \left\lbrace
        \begin{array}{lcl}
        {\bf{F}}=2\omega \nabla \psi\mbox{ for $\Omega$}
        \\
        {\bf{F}}_1=2\omega \nabla \psi_1\mbox{ for $\Omega_1$}
        \end{array}
        \right.
\end{equation}
with $\omega$ being the rotational speed of Earth.\\
Interface velocity potentials $\phi(x,t)$ and $\phi_1(x,t)$ are introduced defined as
\begin{equation}
        \left\lbrace
        \begin{array}{lcl}
        \phi:=\tilde{\varphi}(x,\eta(x,t),t) \mbox{ for $\Omega$}
        \\
        \phi_1:=\tilde{\varphi}_1(x,\eta(x,t),t) \mbox{ for $\Omega_1$}.
        \end{array}
        \right.
\end{equation}
The variable $\xi(x,t),$ introduced in \cite{BenjBrid,BenjBridPart2}, will play the important role of  momentum. It is defined as
\begin{alignat}{2}
\xi:=\rho\phi-\rho_1\phi_1
\end{alignat}

\noindent and belongs to $\mathcal{S}(\mathbb{R})$. Since $\delta\varphi=\delta \tilde{\varphi}$ and $\delta\varphi_1=\delta \tilde{\varphi}_1$, including on $y=\eta(x,t),$ hence the Hamiltonian structure (the Poisson brackets) is the same as  in \cite{Compelli,Compelli2},
that is independent of the constants $\kappa$ and $\kappa_1$ (although the Hamiltonian, in general, depends on $\kappa$ and $\kappa_1$).  The (non-canonical)  equations of motion can be represented in the form \cite{Compelli,Compelli2}

\begin{equation}
\label{EOMsys}
        \left\lbrace
        \begin{array}{lcl}
        \eta_t=\delta_{\xi} H
        \\
        \xi_t=-\delta_{\eta} H+\Gamma  \chi
        \end{array}
        \right.
\end{equation}

\noindent where $H=H(\eta, \xi)$ is the total energy of the system, 
\begin{alignat}{2}
\Gamma:=\rho\gamma-\rho_1\gamma_1+2\omega\big(\rho-\rho_1\big)
\end{alignat}
and
\begin{alignat}{2}
\label{lem2}
\chi(x,t)=- \int_{-\infty}^x\eta_t (x',t)dx'=-\partial_x^{-1}\eta_t
\end{alignat}
is the stream function, evaluated at $y=\eta(x,t),$ (see \cite{Compelli} for details).

The dynamics can be formally cast in the canonical form 
\begin{equation}
        \left\lbrace
        \begin{array}{lcl}
       \eta_t=\delta_{\zeta} H
        \\
        \zeta_t=-\delta_{\eta} H
        \end{array}
        \right.
\end{equation}

\noindent under the transformation (cf. \cite{Wahlen,Compelli,Compelli2})
\begin{alignat}{2}
\label{vartrans}
\xi\rightarrow\zeta=\xi-\frac{\Gamma}{2} \int_{-\infty}^{x} \eta(x',t)\,dx',
\end{alignat}

\noindent which shows that the form 
\begin{equation}
\label{EOMsys1}
        \left\lbrace
        \begin{array}{lcl}
        \eta_t=\delta_{\xi} H
        \\
        \xi_t=-\delta_{\eta} H-\Gamma  \int_{-\infty}^{x} \frac{\delta H}{\delta \xi (x')} dx'
        \end{array}
        \right.
\end{equation}

\noindent  is Hamiltonian too \cite{Compelli,Compelli2}. The condition
$ \int\limits_{\mathbb{R}} \eta(x',t) dx' = 0$ ensures that  $\phantom{*****}$
$\int _{-\infty}^x \eta(x',t)dx' \in \mathcal{S} (\mathbb{R})$  and hence $\zeta(x, t) \in \mathcal{S} (\mathbb{R}).$

The dynamics of the velocity potentials are given explicitly by Euler's equations in terms of the velocity potentials:
\begin{alignat}{2}
 \tilde{\varphi}_{t}+\frac{1}{2}|\nabla \psi|^2  -(\gamma+2\omega)\psi+g y+\frac{p}{\rho}=f(t) \quad \text{in} \quad \Omega
\end{alignat}
and
\begin{alignat}{2}
 \tilde{\varphi}_{1,t}+\frac{1}{2}|\nabla \psi_1|^2 -(\gamma_1+2\omega)\psi_1+g y+\frac{p_1}{\rho_1}=f_1(t) \quad \text{in} \quad \Omega_1
\end{alignat}
for functions $f(t)$ and $f_1(t)$ which we consider as being arbitrary for the moment. 

The time evolution of $\eta(x,t)$ is given by the kinematic boundary condition for the interface $y=\eta(x,t)$ (indicated by the subscript $c$)
\begin{alignat}{2}
\eta_t= (\tilde{\varphi}_{y})_c-\eta_x((\tilde{\varphi}_{x})_c+\gamma\eta+\kappa)= (\tilde{\varphi}_{1,y})_c-\eta_x((\tilde{\varphi}_{1,x})_c+\gamma_1\eta+\kappa_1).
\end{alignat}
At the flat surface on the top $p$ is given by the atmospheric pressure $p_{\text{atm}}$ and so
\begin{alignat}{2}
\bigg[ \tilde{\varphi}_{1,t}+\frac{1}{2}|\nabla \psi_1|^2 -(\gamma_1+2\omega)\psi_1+g h_1+\frac{p_{\text{atm}}}{\rho_1}\bigg]_{y=h_1}=f_1(t).
\end{alignat}
We choose
\begin{alignat}{2}
f_1(t)=gh_1+\frac{p_{\text{atm}}}{\rho_1}
\end{alignat}
therefore
\begin{alignat}{2}
\bigg[  \tilde{\varphi}_{1,t}+\frac{1}{2}|\nabla \psi_1|^2 -(\gamma_1+2\omega)\psi_1\bigg]_{y=h_1}=0.
\end{alignat}

At the interface $y=\eta(x,t)$ we have $p(x,\eta,t)=p_1(x,\eta,t)$ and therefore
\begin{multline}
\rho\Big(({\tilde{\varphi}_{t}})_c+\frac{1}{2}|\nabla \psi|_c^2-(\gamma+2\omega)\chi +g\eta+f(t)\Big)\\=\rho_1\Big(( {\tilde{\varphi}_{1,t}})_c+\frac{1}{2}|\nabla \psi_1|_c^2-(\gamma_1+2\omega)\chi +g\eta+f_1(t)\Big).
\end{multline}

\noindent We choose
\begin{alignat}{2}
\rho f(t)=\rho_1 f_1(t)
\end{alignat}
and so
\begin{alignat}{2}
f(t)= \Big(\frac{\rho_1}{\rho }gh_1+\frac{p_{\text{atm}}}{\rho}\Big).
\end{alignat}
This gives the Bernoulli condition

\begin{multline}
\rho\Big(({\tilde{\varphi}_{t}})_c+\frac{1}{2}|\nabla \psi|_c^2-(\gamma+2\omega)\chi +g\eta\Big)\\=\rho_1\Big(( {\tilde{\varphi}_{1,t}})_c+\frac{1}{2}|\nabla \psi_1|_c^2-(\gamma_1+2\omega)\chi +g\eta\Big).
\end{multline}

\noindent Also, this gives
\begin{alignat}{2}
p=p_{\text{atm}}+\rho_1 gh_1-\rho gy-\rho\Big( \tilde{\varphi}_{t}+\frac{1}{2}|\nabla \psi|^2  -(\gamma+2\omega)\psi  \Big)
\end{alignat}
where $p_{\text{atm}}+\rho_1 gh_1-\rho gy$ is the hydrostatic component and the remaining terms are due to the wave motion
and
\begin{alignat}{2}
p_1=p_{\text{atm}}+\rho_1g (h_1-y)-\rho_1\Big({\tilde{\varphi}_{1,t}}+\frac{1}{2}|\nabla \psi_1|^2-(\gamma_1+2\omega)\psi_1   \Big)
\end{alignat}
where $p_{\text{atm}}+\rho_1g (h_1-y)$ is the hydrostatic component and the remaining terms are due to the wave motion.

The pressure in the body of the fluid can be evaluated from the functions $\tilde{\varphi}(x,y,t),$ $\psi(x,y,t)$ $\tilde{\varphi}_1(x,y,t)$ and $\psi_1(x,y,t).$  It will be demonstrated below that $\tilde{\varphi}(x,y,t)$ and $\tilde{\varphi}_1(x,y,t)$ at $y=\eta(x,t)$ can be recovered from $\xi(x,t).$  In addition, there is an interdependency between  $\tilde{\varphi}, \psi$ and  $\tilde{\varphi}_1, \psi_1$ since $\tilde{\varphi}+i(\psi-\frac{1}{2}\gamma y^2 - \kappa y),$ $\tilde{\varphi}_1+i(\psi_1-\frac{1}{2}\gamma_1 y^2 - \kappa_1 y),$ are analytic functions of the form $z=x+iy.$  Thus, these analytic functions in their respective domains $\Omega$ and $\Omega_1$ can be recovered from their values at the corresponding boundary $\Omega$ or $\Omega_1,$ i.e. from the values at $y=\eta(x,t).$

\section{Evaluation of the Hamiltonian}
The Hamiltonian of the system is given by
\begin{multline}
H=\frac{1}{2}\rho \int\limits_{\mathbb{R}} \jj (u^2+v^2)dy dx'+\frac{1}{2}\rho_1 \int\limits_{\mathbb{R}} \jjj (u_1^2+v_1^2)dy dx'\\+\frac{1}{2}\rho g \int\limits_{\mathbb{R}} \jj y\, dy dx' +\frac{1}{2}\rho_1 g\int\limits_{\mathbb{R}}\jjj y\, dy dx'+ \int\limits_{\mathbb{R}} \mathfrak{h}_0 dx,
\end{multline}
where $\mathfrak{h}_0$ is a constant Hamiltonian density (with zero variations), compensating for any constant terms that arise in the other integrals, so that the overall Hamiltonian density is a function from the class $\mathcal{S}(\mathbb{R})$.
$H$ can be further decomposed, using (\ref{def__stream_phi}), as
\begin{multline}
H=\frac{1}{2}\rho\int\limits_{\mathbb{R}} \jj ((\tilde{\varphi}_x + \gamma y+\kappa)^2+( {\tilde{\varphi}}_{y})^2)dy dx'\\
+\frac{1}{2}\rho_1 \int\limits_{\mathbb{R}} \jjj (( \tilde{\varphi}_{1,x} + \gamma_1 y+\kappa_1)^2+({\tilde{\varphi}}_{1,y})^2)dy dx'\\+\frac{1}{2}\rho g\int\limits_{\mathbb{R}} (\eta^2-h^2) dx +\frac{1}{2}\rho_1 g\int\limits_{\mathbb{R}} (h_1^2-\eta^2) dx + \int\limits_{\mathbb{R}} \mathfrak{h}_0 dx
\end{multline}
which can be written as
\begin{multline}
\label{Ham_1}
H=\frac{1}{2}\rho\int\limits_{\mathbb{R}} \jj ((\tilde{\varphi}_x)^2 + ({\tilde{\varphi}}_{y})^2)dy dx'+\frac{1}{2}\rho_1\int\limits_{\mathbb{R}} \jjj ((\tilde{\varphi}_{1,x})^2 + ({\tilde{\varphi}}_{1,y})^2)dy dx'\\
+\rho\gamma\int\limits_{\mathbb{R}} \jj \tilde{\varphi}_x y\,dy dx'+\rho_1\gamma_1 \int\limits_{\mathbb{R}} \jjj \tilde{\varphi}_{1,x}y\,dy dx'\\
+\rho\kappa\int\limits_{\mathbb{R}} \jj \tilde{\varphi}_x\,dy dx'+\rho_1\kappa_1\int\limits_{\mathbb{R}} \jjj \tilde{\varphi}_{1,x}\,dy dx'\\
+\frac{1}{2}\rho\int\limits_{\mathbb{R}} \jj (\gamma y+ \kappa)^2 dy dx'
+\frac{1}{2}\rho_1\int\limits_{\mathbb{R}} \jjj (\gamma_1 y+ \kappa_1)^2 dy dx'\\ +\frac{1}{2}g(\rho-\rho_1 )\int\limits_{\mathbb{R}} \eta^2 dx+\int\limits_{\mathbb{R}} \mathfrak{h}_1 dx
\end{multline}

\noindent where $\mathfrak{h}_1$ is another appropriate constant.

We introduce the Dirichlet-Neumann operators
\begin{equation}
        \left\lbrace
        \begin{array}{lcl}
        G(\eta)\phi=(\tilde{\varphi}_{{\bf{n}}})\sqrt{1+(\eta_x)^2} \mbox{ for $\Omega$}
        \\
        G_1(\eta)\phi_1=(\tilde{\varphi}_{1,{\bf{n}}_1})\sqrt{1+(\eta_x)^2} \mbox{ for $\Omega_1$}
        \end{array}
        \right.
\end{equation}
where $\varphi_{{\bf{n}}}$ and $\varphi_{1,{\bf{n}}_1}$ are the normal derivatives of the velocity potentials, at the interface, for outward normals ${{\bf{n}}}$ and ${{\bf{n}}_1}$ (noting that ${{\bf{n}}}=-{{\bf{n}}_1}$), see more details in \cite{Craig1,Craig2}.

Using the boundary conditions
\begin{alignat}{2}
        \left\lbrace
        \begin{array}{lcl}
        G(\eta)\phi=-\eta_x(\tilde{\varphi}_x)_c+ (\tilde{\varphi}_y)_c = \eta_t+(\gamma\eta+\kappa)\eta_x,
        \\
        G_1(\eta)\phi_1=\eta_x(\tilde{\varphi}_{1,x})_c-(\tilde{\varphi}_{1,y})_c=-\eta_t-(\gamma_1\eta+\kappa_1)\eta_x
        \end{array}
        \right.
\end{alignat}
we get
\begin{alignat}{2}
G(\eta)\phi+G_1(\eta)\phi_1=\mu
\end{alignat}
where
\begin{alignat}{2}
\mu:=\big((\gamma-\gamma_1)\eta+(\kappa-\kappa_1)\big)\eta_x.
\end{alignat}
Recalling that $\xi=\rho\phi-\rho_1\phi_1$
then 
\begin{alignat}{2}
\rho_1G(\eta)\phi+\rho G_1(\eta)\phi=\rho_1\mu+ G_1(\eta)\xi.
\end{alignat}
We define
\begin{alignat}{2}
\label{B_DEF}
B:=\rho G_1(\eta)+\rho_1 G(\eta)
\end{alignat}
and so
\begin{alignat}{2}\label{phys}
        \left\lbrace
        \begin{array}{lcl}
        \phi=B^{-1}\big(\rho_1\mu+G_1(\eta)\xi \big)
        \\
        \phi_1=B^{-1}\big(\rho\mu-G(\eta)\xi\big )
        \end{array}
        \right.
\end{alignat}
which means
\begin{multline}
\frac{1}{2}\rho\int\limits_{\mathbb{R}} \jj ((\tilde{\varphi}_x)^2 + ({\tilde{\varphi}}_{y})^2)dy dx'+\frac{1}{2}\rho_1\int\limits_{\mathbb{R}} \jjj ((\tilde{\varphi}_{1,x})^2 + ({\tilde{\varphi}}_{1,y})^2)dy dx'\\
=\frac{1}{2}\ii  \xi G(\eta)  B^{-1}G_1(\eta)\xi \,dx
+ \frac{1}{2}\rho\rho_1\ii   \mu   B^{-1}\mu  \,dx.
\end{multline}

\noindent Noting that
$$ \int^{ f_2(x)}_{f_1(x)} F_x(x, y)dy = \left[ \int^{ f_2(x)}_{f_1(x)} F(x, y)dy\right]_x+ F(x, f_1(x))f'_1 (x) - F(x, f_2(x))f'_2(x)$$
and using \eqref{phys} we can write

\begin{multline}
 \rho\kappa \int\limits_{\mathbb{R}} \jj \tilde{\varphi}_x\,dy dx'+\rho_1\kappa_1\int\limits_{\mathbb{R}} \jjj \tilde{\varphi}_{1,x}\,dy dx'\\
\label{Eqtemp1}
=\ii \Big(-\rho\rho_1\kappa B^{-1}\mu-\rho\kappa B^{-1}G_1(\eta)\xi +\rho\rho_1\kappa_1 B^{-1}\mu-\rho_1\kappa_1 B^{-1}G(\eta)\xi \Big) \eta_x dx
\end{multline}
and
\begin{multline}
\rho\gamma\int\limits_{\mathbb{R}} \jj \tilde{\varphi}_x y\,dy dx'+\rho_1\gamma_1 \int\limits_{\mathbb{R}} \jjj \tilde{\varphi}_{1,x}y\,dy dx'\\
\label{Eqtemp2}
=\ii \Big(-\rho\rho_1\gamma B^{-1}\mu-\rho\gamma B^{-1}G_1(\eta)\xi +\rho\rho_1\gamma_1 B^{-1}\mu-\rho_1\gamma_1 B^{-1}G(\eta)\xi \Big)\eta \eta_x\,dx.
\end{multline}

The terms (\ref{Eqtemp1}) and (\ref{Eqtemp2}) combine as
\begin{multline}
-\rho\rho_1\ii \mu B^{-1}\mu \,dx
-\ii (\kappa+\gamma\eta)\xi\eta_x 
\,dx+\rho_1\ii \mu B^{-1}G(\eta)\xi\,dx.
\end{multline}
Also the integrals
\begin{equation}
\frac{1}{2}\rho\int\limits_{\mathbb{R}} \jj (\gamma y+ \kappa)^2 dy dx'
+\frac{1}{2}\rho_1\int\limits_{\mathbb{R}} \jjj (\gamma_1 y+ \kappa_1)^2 dy dx'
\end{equation}
\noindent produce the terms
\begin{equation}
\frac{\rho}{6\gamma}\ii (\gamma \eta+\kappa)^3dx-\frac{\rho_1}{6\gamma_1}\ii (\gamma_1 \eta+\kappa_1)^3dx
\end{equation} 

\noindent noting that e.g. $\ii (\gamma \eta+\kappa)^3dx$ can be properly re-normalised as $$\ii [(\gamma \eta+\kappa)^3-\kappa^3]dx$$ 

\noindent by introducing a constant density integral $\ii \kappa^3 dx$ having in mind that $\eta \in \mathcal{S}(\mathbb{R})$ and that the constant densities are compensated by $\mathfrak{h}_1$. The Hamiltonian can hence be written as
\begin{multline}
\label{Main_Ham}
H(\eta,\xi)=\frac{1}{2}\ii \xi G(\eta) B^{-1}G_1(\eta)\xi \,dx
- \frac{1}{2}\rho\rho_1\ii   \mu   B^{-1}\mu  \,dx\\
-\ii (\kappa+\gamma\eta)\xi\eta_x \,dx+\rho_1\ii \mu B^{-1}G(\eta)\xi\,dx\\
+\frac{\rho}{6\gamma}\ii (\gamma \eta+\kappa)^3dx-\frac{\rho_1}{6\gamma_1}\ii (\gamma_1 \eta+\kappa_1)^3dx +\frac{1}{2}g(\rho-\rho_1 )\int\limits_{\mathbb{R}} \eta^2 dx+\int\limits_{\mathbb{R}}  \mathfrak{h}_1 dx,
\end{multline} where

$$ \mathfrak{h}_1=\frac{\rho_1\kappa_1^3}{6\gamma_1}- \frac{\rho\kappa^3}{6\gamma}.$$

Considering the case where $\gamma_1=\gamma$ and $\kappa_1=\kappa$ then $\mu=0$ and the Hamiltonian can be written as
\begin{multline}
\label{HamPlisk}
H(\eta,\xi)=\frac{1}{2}\ii  (G(\eta)\xi  B^{-1}G_1(\eta))\xi \,dx
-\ii (\kappa+\gamma\eta)\xi\eta_x \,dx\\
+\frac{\rho-\rho_1}{6\gamma}\ii [(\gamma \eta+\kappa)^3-\kappa^3]dx +\frac{1}{2}g(\rho-\rho_1 )\int\limits_{\mathbb{R}} \eta^2 dx
\end{multline}
thus recovering the Hamiltonian determined in \cite{CompelliIvanov2}.

In the case where $\gamma_1\neq\gamma$ and $\kappa_1=\kappa=0$ then $\mu=(\gamma-\gamma_1)\eta\eta_x$ and the Hamiltonian can be written as
\begin{multline}
\label{HamMonat}
H(\eta,\xi)=\frac{1}{2}\ii \xi G(\eta) B^{-1}G_1(\eta)\xi \,dx
+\rho_1 (\gamma-\gamma_1)\ii\eta\eta_x B^{-1}G(\eta)\xi\,dx\\
- \frac{1}{2}\rho\rho_1(\gamma-\gamma_1)^2\ii  \eta\eta_x B^{-1}\eta\eta_x  \,dx
-\gamma\ii \xi\eta\eta_x \,dx\\
+\frac{1}{6}(\rho\gamma^2-\rho_1\gamma_1 ^2)\ii \eta^3dx +\frac{1}{2}g(\rho-\rho_1 )\int\limits_{\mathbb{R}} \eta^2 dx
\end{multline}
which, by an application of the definition of $B$ from (\ref{B_DEF}), recovers the result in \cite{Compelli2} (noting that there is a sign difference is the second and fourth terms due to the different stream function convention used in \cite{Compelli2}).

\section{Scales and small amplitude expansion}
\label{sect:Scalesandexpansion}

In order to describe the scales, let us introduce non-dimensional variables (without bars) related to the dimensional (barred) as follows:

\begin{equation} \label{nondim}
\begin{split}
\bar{t}=&\frac{h_1}{\sqrt{gh_1}}t, \qquad \bar{x}=h_1x, \qquad \bar{y}=h_1y, \qquad \bar{\eta}=a\eta, \qquad  \bar{u}=\sqrt{gh_1}u, \\ \bar{v}&=\sqrt{gh_1} v, \quad \bar{\kappa}=\sqrt{gh_1}\kappa, \quad   \bar{\gamma}=\frac{\sqrt{gh_1}}{h_1}\gamma, \quad  \bar{\gamma}_1=\frac{\sqrt{gh_1}}{h_1}\gamma_1,  \quad \varepsilon=\frac{a}{h_1}.
\end{split}
\end{equation}

The constant $a$ represents the average amplitude of the waves $\eta(x,t)$ under consideration and $\varepsilon$ is a small parameter which will be used to separate the order of the terms in the model. From $\bar{v}=\bar{\eta}_{\bar{t}}+\bar{u}\bar{\eta}_{\bar{x}}$ it follows that  \begin{equation} \label{nondim3}
v=\varepsilon(\eta_t+u\eta_x).
\end{equation} Therefore, if $\eta_t=\mathcal{O}(1)$ then $v=\mathcal{O}(\varepsilon)$ and thus the dimensional expression with $v=\mathcal{O}(1)$ (and similar for $v_1$) should be 

\begin{equation} \label{nondim4}
\bar{v}=\varepsilon \sqrt{gh_1} v.
\end{equation}

Since $v$ is a $y$-derivative of the velocity potential, and with the adopted definitions $\bar{\varphi}=\varepsilon h_1 \sqrt{gh_1}\varphi $ etc., then

\begin{equation} \label{nondim5}
\bar{\xi}=\varepsilon \rho h_1 \sqrt{gh_1} \xi.
\end{equation}  The other scales do not change - their dominant parts are the vorticity  and current components of order 1; only the 'wave' component (which is the $x-$derivative of  $\varphi $) is of order $\varepsilon$.
The Dirichlet-Neumann operators have the following structure   

\begin{equation} \label{DN}
\bar{G}=\bar{G}^{(0)} +\bar{ G}^{(1)}+\bar{G}^{(2)}+ \ldots
\end{equation} 
where $\bar{G}^{(n)}\sim \bar{\eta}^n \partial_{\bar{x}}^{n+1}$, i.e. $\bar{G}^{(n)}=\frac{\varepsilon^n}{h_1}G^{(n)}$, and similarly for $G_{1}$, and so  


\begin{alignat}{2}
\label{DN_G}
\bar{G}(\bar{\eta})&=\bar{D}\tanh(h \bar{D})+\bar{D} \bar{\eta}  \bar{D} -\bar{D}\tanh(h \bar{D}) \bar{\eta} \bar{D}\tanh(h \bar{D})+\mathcal{O}(\varepsilon^2)\\
\label{DN_G1}
\bar{G}_1(\bar{\eta})&=\bar{D}\tanh(h_1 \bar{D})-\bar{D} \bar{\eta} \bar{D} +\bar{D}\tanh(h_1 \bar{D}) \bar{\eta} \bar{D}\tanh(h_1 \bar{D})+\mathcal{O}(\varepsilon^2)
\end{alignat}
where $\bar{D}=-i\partial/\partial \bar{x}$.

The operator $\bar{B}$, which is a function of Dirichlet-Neumann operators, can hence be expressed as
\begin{alignat}{2}
\bar{B}=\rho_1\sum_{j=0}^{\infty} \bar{G}_{j}(\bar{\eta})+\rho\sum_{j=0}^{\infty} \bar{G}_{1j}(\bar{\eta}).
\end{alignat}

With this scaling, and ignoring the linear terms (whose average is 0), the Hamiltonian can be expanded as 
\begin{equation} \label{Hexp}
\bar{H}= \rho g h_1^3\left( \varepsilon^2 H^{(2)} +\varepsilon^3 H^{(3)}+ \ldots \right).
\end{equation}

Recalling the Dirichlet-Neumann operators in (\ref{DN_G}) and (\ref{DN_G1}), noting that $$\eta_x [B^{-1}]^{(0)}\eta_x=-\eta D^2 [B^{-1}]^{(0)}\eta,$$ this gives
\begin{multline}
H^{(2)}=\frac{1}{2}\ii \xi \frac{D\tanh(h D)\tanh(h_1 D)}{\rho \tanh(h_1 D)+\rho_1 \tanh(h D)}\xi \,dx\\
+ \frac{1}{2}\rho\rho_1 (\kappa-\kappa_1)^2\ii \eta \frac{D}{\rho \tanh(h_1 D)+\rho_1  \tanh(h D)}\eta \,dx-\kappa\ii \xi\eta_x \,dx\\
+\rho_1(\kappa-\kappa_1)\ii \eta_x \frac{\tanh(h D)}{\rho \tanh(h_1 D)+\rho_1 \tanh(h D)}\xi\,dx
+\frac{1}{2}A_1\ii \eta^2 \,dx
\end{multline}
where $$A_1=\rho\gamma\kappa-\rho_1 \gamma_1 \kappa_1+g(\rho-\rho_1 ).$$
This gives the linear equations
\begin{multline}
\eta_t+\kappa \eta_x=\frac{D\tanh(h D)\tanh(h_1 D)}{\rho \tanh(h_1 D)+\rho_1 \tanh(h D)}\xi\\
+ \frac{\rho_1(\kappa-\kappa_1)\tanh(h D)}{\rho \tanh(h_1 D)+\rho_1 \tanh(h D)}\eta_x
\end{multline}
and 
\begin{multline}
\xi_t+ \kappa \xi_x+\Gamma\int_{-\infty}^x\eta_t\,dx'= \frac{\rho_1(\kappa-\kappa_1)\tanh(h D)}{\rho \tanh(h_1 D)+\rho_1 \tanh(h D)}\xi_x\\
-  \frac{\rho\rho_1 (\kappa-\kappa_1)^2 D}{\rho \tanh(h_1 D)+\rho_1  \tanh(h D)}\eta -A_1\eta.
\end{multline}

Looking for a solution which is a superposition of  sine and cosine waves, we represent $\eta$ and $\xi$ as
\begin{equation}
        \left\lbrace
        \begin{array}{lcl}
        \eta(x,t)=\eta_0e^{-i(\Omega(k) t-kx)}
        \\
        \xi(x,t)=\xi_0e^{-i(\Omega(k) t-kx)}
        \end{array}
        \right.
\end{equation}
where $k$ is the wave number and $\Omega(k)$ is the angular frequency. The wave speed $c$ is given by
\begin{alignat}{2}
c(k)=\frac{\Omega(k)}{k}.
\end{alignat}
 
\noindent Introducing the functions
\begin{equation}
        \left\lbrace
        \begin{array}{lcl}
        T(k)=\tanh(k h)
        \\
        T_1(k)=\tanh(k h_1)
        \end{array}
        \right.
\end{equation}
one can obtain the following quadratic equation for the wave speed
\begin{multline}
(c-\kappa)^2+\frac{2(\kappa-\kappa_1)\rho_1 k T+\Gamma T T_1 }{k(\rho_1 T+\rho T_1)}(c-\kappa)-\frac{(A_1-\kappa\Gamma)T T_1}{k(\rho_1 T+\rho T_1)}\\+\frac{(\kappa-\kappa_1)^2  \rho_1 T (\rho_1 T-\rho T_1)}{(\rho_1 T+\rho T_1)^2}=0.
\end{multline}
For $\kappa=\kappa_1$ we have 
$$A_1-\kappa \Gamma = g(\rho-\rho_1 )- 2\kappa \omega(\rho-\rho_1),$$ which is a quantity usually close to $ g(\rho-\rho_1 )$ and thus positive (see the discussion below), then the quadratic equation becomes

\begin{alignat}{2}
(c-\kappa)^2+\frac{\Gamma T T_1 }{k(\rho_1 T+\rho T_1)}(c-\kappa)-\frac{(A_1-\kappa \Gamma)T T_1 }{k(\rho_1 T+\rho T_1)}=0
\end{alignat}
with solutions
\begin{alignat}{2}
c_{\pm}=\kappa+\frac{1}{2}\bigg[-f_1\pm\sqrt{f_1^2+4 f_2}\bigg]
\end{alignat}
where
\begin{alignat}{2}
f_1(k)= \frac{\Gamma \tanh (k h ) \tanh (k h_1 ) }{k(\rho_1 \tanh (k h )+\rho \tanh (k h_1 ))}
\end{alignat}
and
\begin{alignat}{2}
f_2(k)= \frac{(A_1-\kappa \Gamma) \tanh (k h ) \tanh (k h_1 ) }{k(\rho_1 \tanh (k h )+\rho \tanh (k h_1 ))}
\end{alignat}
corresponding to right moving ($c_{+}>\kappa$) and left moving ($c_{-}<\kappa$) waves with respect to a {\it moving} observer, moving with the a velocity $\kappa.$ The reference frame for the {\it moving} observer is obtained by a Galilean transformation $X\to x-\kappa t,$  $T \to t,$
$$ \partial_T \rightarrow \partial_t + \kappa \partial_x, \qquad \partial_X \to \partial_x$$
giving
\begin{alignat}{2}
\eta_T=\frac{DT(D) T_1(D)}{\rho T_1(D)+\rho_1 T(D)}\xi
+\frac{\rho_1(\kappa-\kappa_1) T(D)}{\rho T_1(D)+\rho_1 T(D)}\eta_x
\end{alignat}
and 
\begin{multline}
\xi_T+\Gamma \partial_x^{-1}\eta_T=(\Gamma\kappa-A_1)\eta+\frac{\rho_1(\kappa-\kappa_1) T(D)}{\rho T_1(D)+\rho_1 T(D)}\xi_x
+  \frac{\rho\rho_1 (\kappa-\kappa_1)^2D}{\rho T_1+\rho_1 T}\eta
.
\end{multline}
Consider the term $(\Gamma\kappa-A_1)\eta$. This evaluates to
\begin{alignat}{2}
\rho_1\gamma_1(\kappa_1-\kappa)-(\rho-\rho_1)g+2\kappa(\rho-\rho_1)\omega.
\end{alignat}
 
Note that the term $2\kappa(\rho-\rho_1)\omega$ is the only term in the linearised equation that depends on $\kappa$ but does not depend on the relative velocity $\kappa-\kappa_1$. All other terms depend only on the difference $\kappa-\kappa_1$. This is a consequence of the presence of the Coriolis term representing non-inertial forces in our frame of reference. When $\omega=0$ the equations, of course, depend only on the relative difference $\kappa-\kappa_1$. 

If $\kappa-\kappa_1 \ne 0$ there is a jump in the velocity component, tangent to the surface of the internal wave $y=\eta(x,t),$ i.e. there is a vortex sheet, see the discussion in \cite{ConstIvMartin}. While such a situation is compatible with the inviscid Euler's equations, in practical situations, where viscosity is always present, such jumps do not occur (otherwise there will be a vortex sheet between the two media). For this reason in our further considerations we take $\kappa_1=\kappa.$ This corresponds to the case where there is no vortex sheet at the boundary between the two layers at $y=\eta(x,t)$.

Recalling the Hamiltonian in (\ref{Main_Ham}) where, in this case, $\mu=(\gamma-\gamma_1)\eta\eta_x$
\begin{multline}
H(\eta,\xi)=\frac{1}{2}\ii \xi G(\eta) B^{-1}G_1(\eta)\xi \,dx
- \frac{1}{2}\rho\rho_1(\gamma-\gamma_1)^2\ii   \eta\eta_x   B^{-1}\eta\eta_x \,dx\\
-\ii (\kappa+\gamma\eta)\xi\eta_x \,dx+\rho_1(\gamma-\gamma_1)\ii\eta\eta_x B^{-1}G(\eta)\xi\,dx\\
+\frac{1}{2}[(\rho-\rho_1)g+(\rho\gamma-\rho_1\gamma_1)\kappa]\int\limits_{\mathbb{R}} \eta^2 dx+\frac{1}{6}(\rho\gamma^2-\rho_1\gamma_1^2)\ii  \eta^3dx
\end{multline}
noting re-normalisation to exclude any constant terms. 

\section{Long waves approximation}

We will study the equations under the additional approximation that the wavelengths $L$ are much bigger than $h$ and $h_1$. Since $$\bar{L}=h_1 L \Rightarrow \frac{1}{L}=\frac{h_1}{\bar{L}}=\delta $$ Thus for the wave number $k=2\pi/L=2\pi \delta$ we have $k=\mathcal{O}(\delta) $. We further assume that $\delta^2=\mathcal{O}(\varepsilon).$
Recall that the operator $D$ has an eigenvalue $k$, thus we shall keep in mind that $D=\mathcal{O}(\delta). $ Moreover the $x$-derivative of the velocity potentials do not get an extra factor of $\delta$ since $\bar{v}$ of order $\varepsilon$ remains unchanged. In other words the ``wave'' component of $u$ is $\tilde{\varphi}_x$ and is of order $\varepsilon\sim \delta^2,$ hence $\tilde{\varphi}_x \sim \delta$ and  $\xi \sim \delta.$

We write now all scale factors explicitly, and keep all remaining quantities non-dimensional and of order one. The Dirichlet-Neumann operators can hence be represented as
\begin{multline}
G(\eta)=\delta\Big( D\tanh(\delta h D)\Big)\\+\varepsilon\delta^2\Big(  D   \eta D-  D\tanh(\delta h D)  \eta  D\tanh(\delta h D)\Big)+\mathcal{O}(\varepsilon^2\delta^4)
\end{multline}
and
\begin{multline}
G_1(\eta)=\delta\Big( D \tanh(\delta h_1 D)\Big)\\- \varepsilon\delta^2\Big(D\eta  D -  D \tanh(\delta h_1 D)  \eta  D \tanh(\delta h_1 D)\Big)+\mathcal{O}(\varepsilon^2\delta^4).
\end{multline}

Expanding the hyperbolic tangent functions the Dirichlet-Neumann operators can be represented as
\begin{multline}
G(\eta)=\delta^2 \bigg( h D^2+\varepsilon D \eta D\bigg)-\delta^4\bigg(\frac{1}{3} h^3 D^4+\varepsilon  h^2 D^2 \eta D^2\bigg )\\ +\delta^6\bigg(\frac{2}{15}h^5 D^6\bigg) 
+\mathcal{O}(\delta^8,\varepsilon\delta^6,\varepsilon^2\delta^4)
\end{multline}
and
\begin{multline}
G_1(\eta)=\delta^2\bigg( h_1 D^2-\varepsilon D \eta D\bigg)+\delta^4\bigg(-\frac{1}{3}h_1^3 D^4+\varepsilon  h_1^2 D^2 \eta D^2 \bigg)\\+\delta^6\bigg(\frac{2}{15}h_1^5 D^6 \bigg)
+\mathcal{O}(\delta^8,\varepsilon\delta^6,\varepsilon^2\delta^4).
\end{multline}

 $B$, the non-dimensional form of $\bar{B}$, is given as:
\begin{multline}
B=\delta^2\bigg( (\rho_1 h+\rho h_1)D^2
+\varepsilon(\rho_1 -\rho)D \eta D\bigg)\\
-\delta^4\bigg( \frac{1}{3}(\rho_1  h^3+\rho h_1^3)D^4
+\varepsilon(\rho_1 h^2 -\rho h_1^2)D^2 \eta D^2\bigg)\\
+\delta^6\bigg( \frac{2}{15}(\rho_1  h^5+\rho  h_1^5)D^6
\bigg)+\mathcal{O}(\delta^8,\varepsilon\delta^6,\varepsilon^2\delta^4).
\end{multline}
This can be written as
\begin{multline}
{B}=\delta^2 (\rho_1 h+\rho h_1)D\Bigg\{1+\varepsilon\bigg(\frac{\rho_1 -\rho}{\rho_1 h+\rho h_1}\bigg) \eta\\-\delta^2\Bigg(\frac{1}{3}\bigg(\frac{\rho_1  h^3+\rho h_1^3}{\rho_1 h+\rho h_1}\bigg)D^2 +\varepsilon \bigg(\frac{\rho_1 h^2 -\rho h_1^2}{\rho_1 h+\rho h_1}\bigg)D \eta D\Bigg)\\+\delta^4\Bigg(\frac{2}{15}\bigg(\frac{\rho_1  h^5+\rho  h_1^5}{\rho_1 h+\rho h_1}\bigg)D^4\Bigg)+\mathcal{O}(\delta^6,\varepsilon\delta^4,\varepsilon^2\delta^2) \Bigg\}D
\end{multline}
and so
\begin{multline}
{B}^{-1}=\frac{1}{\delta^2 (\rho_1 h+\rho h_1)}D^{-1}\Bigg\{1+\varepsilon\bigg(\frac{\rho_1 -\rho}{\rho_1 h+\rho h_1}\bigg) \eta \\-\delta^2\Bigg(\frac{1}{3}\bigg(\frac{\rho_1  h^3+\rho h_1^3}{\rho_1 h+\rho h_1}\bigg)D^2 +\varepsilon \bigg(\frac{\rho_1 h^2 -\rho h_1^2}{\rho_1 h+\rho h_1}\bigg)D \eta D\Bigg)\\+\delta^4\Bigg(\frac{2}{15}\bigg(\frac{\rho_1  h^5+\rho  h_1^5}{\rho_1 h+\rho h_1}\bigg)D^4\Bigg)+\mathcal{O}(\delta^6,\varepsilon\delta^4,\varepsilon^2\delta^2)\Bigg\}^{-1}D^{-1}.
\end{multline}
Using the expansion $(1+x)^{-1}=1-x+x^2-x^3+\mathcal{O}(x^4)$ we obtain
\begin{multline}
{B}^{-1}=\frac{1}{\delta^2 (\rho_1 h+\rho h_1)}D^{-1}\Bigg\{1-\varepsilon\bigg(\frac{\rho_1 -\rho}{\rho_1 h+\rho h_1}\bigg) \eta+\varepsilon^2\bigg(\frac{(\rho_1 -\rho )^2}{(\rho_1 h+\rho h_1)^2}\bigg) \eta^2 \\
+\delta^2\Bigg(\frac{1}{3}\bigg(\frac{\rho_1  h^3+\rho h_1^3}{\rho_1 h+\rho h_1}\bigg)D^2-\frac{1}{3}\varepsilon \bigg(\frac{(\rho_1 -\rho )(\rho_1  h^3+\rho h_1^3)}{(\rho_1 h+\rho h_1)^2}\bigg) \eta D^2\\ -\frac{1}{3} \varepsilon\bigg(\frac{(\rho_1 -\rho )(\rho_1  h^3+\rho h_1^3)}{(\rho_1 h+\rho h_1)^2}\bigg)D^2 \eta +\varepsilon \bigg(\frac{\rho_1 h^2 -\rho h_1^2}{\rho_1 h+\rho h_1}\bigg)D \eta D\Bigg)\\
-\delta^4\Bigg(\frac{2}{15}\bigg(\frac{\rho_1  h^5+\rho  h_1^5}{\rho_1 h+\rho h_1}\bigg)D^4  -\frac{1}{9}\bigg(\frac{(\rho_1  h^3+\rho h_1^3)^2}{(\rho_1 h+\rho h_1)^2}\bigg)D^4\Bigg)\Bigg\}D^{-1}+\mathcal{O}(\delta^6,\varepsilon\delta^4,\varepsilon^2\delta^2,\varepsilon^3).
\end{multline}

The Hamiltonian expanded to terms of orders up to $\delta^{8}$ and $\varepsilon^{4}$ has the  form

\begin{multline}
H(\eta,\xi)=\frac{\delta^4}{2}\ii \xi   
 \Bigg(
\frac{h h_1}{\rho_1 h+\rho h_1}{D}^2
+\varepsilon\frac{\rho h_1^2-\rho_1 h^2}{(\rho_1 h+\rho h_1)^2}
D \eta  D -\varepsilon^2  \frac{\rho \rho_1(h+h_1)^2}{(\rho_1 h+\rho h_1)^3}D \eta^2  D\Bigg)\xi  dx\\
+\frac{\delta^6}{2}\ii  \xi   
 \Bigg(-\frac{h^2 h_1^2(\rho_1   h_1+\rho h)}{3(\rho_1 h+\rho h_1)^2}{D}^4
-\varepsilon \frac{h^2 h_1^2(\rho-\rho_1)}{(\rho_1 h+\rho h_1)^2} {D}^2\eta  {D}^2\\
+ \varepsilon \frac{\rho \rho_1 h h_1(h- h_1)(h+ h_1)^2}{3(\rho_1 h+\rho h_1)^3}(D \eta D^3+D^3 \eta D)
\Bigg)\xi  dx\\
+\frac{\delta^8}{2}\ii \xi   
 \frac{ h^2 h_1^2\left( \rho \rho_1 (h^2-h_1^2)^2+6 h h_1  (\rho h+\rho_1 h_1)^2\right)}{45(\rho_1 h+\rho h_1)^3} D^6
\xi  dx\\
-\varepsilon\delta^2 \kappa\ii \xi\eta_x \,dx+\varepsilon^2\delta^2\frac{\rho_1\gamma_1 h + \rho \gamma h_1}{2(\rho_1 h+\rho h_1)}\ii\eta^2 \xi_x\,dx 
\\
+\varepsilon^3\delta^2\frac{\rho \rho_1 (\gamma-\gamma_1)(h+h_1)}{2(\rho_1 h+\rho h_1)^2}\ii\eta^{3}\xi_x \,dx 
+\varepsilon^2 \delta^4\frac{\rho \rho_1h h_1(\gamma-\gamma_1)(h^2-h_1^2)}{6(\rho_1 h+\rho h_1)^2}\ii\eta^{2}\xi_{xxx}\,dx\\
+\frac{\varepsilon^2}{2}[g(\rho-\rho_1 )+(\rho\gamma-\rho_1\gamma_1)\kappa]\ii  \eta^2 dx +\frac{\varepsilon^3}{6}(\rho\gamma^2-\rho_1\gamma_1^2)\ii  \eta^3dx \\
- \frac{\varepsilon^4}{2}\frac{\rho\rho_1(\gamma-\gamma_1)^2}{\rho_1 h+\rho h_1}\ii   \frac{\eta^4}{4} \,dx.
\end{multline}


\section{ KdV approximation}




 We will keep track only of the scale variables $\varepsilon, \delta$ and not of the other dimensional factors. The Hamiltonian has the following expansion to order $\delta^6$:

\begin{multline} \label{H}
H(\eta,\xi)=\frac{1}{2}\delta^4 \ii \xi D\left( \alpha_1 + \delta^2 (\alpha_3 \eta - \alpha_2 D^2) \right)D \xi dx   +\delta^4\alpha_5 \int\limits_{\mathbb{R}} \frac{\eta^2}{2} dx \\
-\delta^4 \kappa \int\limits_{\mathbb{R}} \eta_x \xi dx -\delta^6 \alpha_4 \int\limits_{\mathbb{R}} \eta \eta_x \xi dx      +\delta^6 \alpha_6\ii  \frac{\eta^3}{6}dx
\end{multline}


where 

\begin{equation} \begin{split} \alpha_1&=\frac{h h_1}{\rho_1 h+\rho h_1}, \qquad \alpha_2= \frac{h^2h_1^2(\rho h+\rho_1 h_1)}{3(\rho_1 h+\rho h_1)^2}, \qquad \alpha_3= \frac{\rho h_1^2-\rho_1 h^2}{(\rho_1 h+\rho h_1)^2}, \\
\alpha_4&=\frac{\gamma_1\rho_1 h+ \gamma \rho h_1}{\rho_1 h+\rho h_1}, \quad \alpha_5= g(\rho-\rho_1)+(\rho \gamma -\rho_1 \gamma_1)\kappa, \quad \alpha_6=\rho\gamma^2-\rho_1 \gamma_1^2.
\end{split}\end{equation}

The Hamiltonian equations \eqref{EOMsys} for the Hamiltonian \eqref{H} in terms of $\eta$ and $\tilde{u}=\xi_x$ are
\begin{equation}\label{BA0}
\begin{split}
&\eta_T + \alpha_1 \tilde{u}_x + \delta^2 \alpha_2 \tilde{u}_{xxx} + \delta^2 (\alpha_3 (\eta \tilde{u})_x+ \alpha_4 \eta \eta_x)=0 \\
&\tilde{u}_T+\Gamma \eta_T +(\rho-\rho_1)(g-2\omega \kappa) \eta_x + \delta^2 (\alpha_3 \tilde{u}\tilde{u}_x + \alpha_4(\tilde{u}\eta)_x + \alpha_6 \eta \eta_x)=0,
\end{split}
\end{equation}

\noindent where for convenience $\partial_T=\partial_t + \kappa \partial_x,$  that  is a Galilean change of coordinates. Since $\omega=7.3\times 10^{-5}$ rad/s, $\kappa \sim 1 $ m/s, then $g\gg 2 \omega \kappa$ and the $2 \omega \kappa$ term will be neglected. One can also exclude $\eta_T$ from the second equation, which leads to the system
\begin{equation}\label{BA}
\begin{split}
\eta_T + \alpha_1 \tilde{u}_x +& \delta^2 \alpha_2 \tilde{u}_{xxx} + \delta^2 (\alpha_3 (\eta \tilde{u})_x+ \alpha_4 \eta \eta_x)=0 \\
\tilde{u}_T-\alpha_1\Gamma \tilde{u}_x  +& (\rho-\rho_1)g\eta_x - \delta^2 \Gamma \alpha_2\tilde{u}_{xxx} \\&+ \delta^2 \left(\alpha_3\tilde{u}\tilde{u} _x + (\alpha_4-\Gamma \alpha_3)(\tilde{u}\eta)_x + (\alpha_6-\Gamma \alpha_4) \eta \eta_x \right )=0.
\end{split}
\end{equation}

\noindent The linearised equations produce wave speeds 
\begin{equation}
c=\frac{1}{2}\left(-\alpha_1 \Gamma \pm \sqrt{\alpha_1 \Gamma^2 +4 \alpha_1(\rho-\rho_1)g}\right)
\end{equation}
for an observer, moving with the flow, i.e. there are left- ($-$ sign) and right-running ($+$ sign) waves. For a stationary observer the velocities are $c+\kappa.$ Moreover, in the leading approximation,  
\begin{equation}
\eta=\frac{\alpha_1}{c}\tilde{u}\qquad\mbox{ and}\qquad \tilde{u}=\frac{c}{\alpha_1} \eta.
\end{equation}

\noindent One can look for a relation between $\eta$ and $u$ at the next order of approximation:
\begin{equation} \tilde{u}=\frac{c}{\alpha_1} \eta + \delta^2 \sigma \eta_{xx} + \delta^2 \mu \eta^2 \label{JT}\end{equation} for some constants $\sigma$ and $\mu$ \cite{J02,J03,J03a}. Then one can express $u$ via $\eta$ in both equations  \eqref{BA}. The condition for the two equations to coincide with terms to order $\delta^2$ leads to the determination of $\sigma$ and $\mu$: 
\begin{equation} \label{sigma}
\sigma=-\frac{c\alpha_2(c+\Gamma \alpha_1)}{\alpha_1^2(2c+\Gamma \alpha_1)}\qquad\mbox{and} \qquad \mu=\frac{\alpha_1 \alpha_4(c-\Gamma \alpha_1)-c\alpha_3(c+2\Gamma \alpha_1) +\alpha_1^2 \alpha_6}{2\alpha_1^2(2c+\Gamma \alpha_1)}.
\end{equation} Thus $u$ can be expressed with $\eta$ to order $\delta^2$ via  \eqref{JT}, \eqref{sigma} and $\eta$ satisfies the KdV equation \cite{KdV}, that represents the coinciding terms of \eqref{BA}:

\begin{equation} \label{KdV} 
\eta_T+c\eta_x+\delta^2\frac{c^2 \alpha_2}{\alpha_1(2c+\Gamma \alpha_1) }\eta_{xxx}+\delta^2\frac{3c^2\alpha_3+3c\alpha_1\alpha_4+\alpha_1^2\alpha_6 \alpha_2}{\alpha_1(2c+\Gamma \alpha_1) }\eta \eta_x=0.
\end{equation}

In the case when all vorticities are zero this simplifies to 
\begin{equation} \label{KdV0} 
\eta_T+c\eta_x+\delta^2\frac{c \alpha_2}{2\alpha_1 }\eta_{xxx}+\delta^2\frac{3c\alpha_3}{2\alpha_1}\eta \eta_x=0.
\end{equation}

The KdV approximation for an internal wave coupled to a free surface is derived in \cite{CI}. 

The KdV equation represents a balance between a nonlinearity term $\eta \eta_x$, and dispersion term $\eta_{xxx}.$ Reintroducing $\varepsilon$ and $\delta,$ it is clear that these terms are scaled like  $\varepsilon \eta \eta_x$ and  $\delta^2\eta_{xxx},$  so that when $\varepsilon \sim \delta^2$ the interplay between nonlinearity and dispersion is producing smooth and stable in time soliton solutions. However, there are various geophysical scales and many other situations are possible, including $\delta \sim \varepsilon^2.$ In such case $\delta^2 \sim \varepsilon^4 \ll \varepsilon$ and instead of a KdV equation the relevant model is the dispersionless Burgers equation ($\partial_{\tau}=\partial_T+c \partial_x$)

\begin{equation} \label{B} 
\eta_{\tau}+\varepsilon\frac{3c^2\alpha_3+3c\alpha_1\alpha_4+\alpha_1^2\alpha_6 \alpha_2}{\alpha_1(2c+\Gamma \alpha_1) }\eta \eta_x=0.
\end{equation}

Such an equation does not support globally smooth solutions, i.e. the solutions always form a vertical slope and break. Such wave-breaking phenomenon is well known for internal waves in the ocean. This is a mechanism that causes mixing in the deep ocean, with implications for biological productivity and sediment transport \cite{L}.

There are other integrable systems which provide an approximation of the equations in the Boussinesq regime, such as the Kaup-Boussinesq system
investigated firstly by D.J. Kaup \cite{K76}. Under the usual scaling the KB system is asymptotically equivalent to the KdV regime. However, from the point of view of the soliton theory, the KB system is by far more complex and sophisticated than the KdV equation. The soliton solutions of the KB system are not yet completely studied, although there are some special soliton solutions obtained in  \cite{K76,SS,IL12}. Other 2-component integrable systems, that can match the model equations up to order $\delta^2$, are the 2-component Camassa-Holm system and the Zakharov-Ito system \cite{CI08,I09,HI,FGL,EHKL}.

\section{Discussion and conclusions}

We studied a two-media system of liquids with different densities, separated by an free internal surface. Both the top and the bottom of the system are considered horizontal and flat. Underlying currents with constant vorticities are present in both layers. The internal waves are formed at the interface between the layers due to gravity and Coriolis forces.  
The model is aimed at geophysical applications, where a typical configuration is the one of a thin shallow layer of warm and less dense water over a much deeper layer of cold denser water. The two layers are separated by a sharp thermocline/pycnocline, where the internal waves are formed. The governing equations are written in a canonical Hamiltonian form, which gives rise to a systematic approach for possible approximations. In particular, small amplitude and long-wave regimes are studied. There are various geophysical scales, allowing for smooth solitons at the KdV regime as well as breaking waves in the very large wavelengths, when the equations are asymptotically equivalent to the dispersionless Burgers equation. A possible limitation of the model is the assumption of a flat surface, which apparently changes the nature of the internal waves. In the case of a free surface, even in the case of very small amplitudes, the internal wave (in a linear approximation) is usually coupled to the surface wave. This has an impact on the possible propagation speeds \cite{ConstIvMartin,CI}. Other asymptotic regimes, e.g. related to the Nonlinear Schr\"odinger equation \cite{TKM} remain to be studied.

\section*{Acknowledgements}
A.C. is funded by the Fiosraigh Scholarship Programme of Dublin Institute
of Technology. R.I. acknowledges Seed funding grant support from Dublin
Institute of Technology for a project in association with the Environmental
Sustainability and Health Institute, Dublin. The authors are grateful to
Prof. Adrian Constantin for many valuable discussions and to two anonymous
referees for several very helpful suggestions and comments.

\end{document}